\documentstyle[aps,prl,multicol,graphicx,epsfig]{revtex}

\begin{document}
\title{Electrical contacts to nanotubes and nanowires: why size matters}
\author{Fran\c{c}ois L\'{e}onard$^{\ast }$ and A. A. Talin}
\address{Sandia National Laboratories, Livermore, California 94551}
\date{\today }
\maketitle

\begin{abstract}
Electrical contacts to semiconductors play a key role in electronics. For
nanoscale electronic devices, particularly those employing novel
low-dimensionality materials, contacts are expected to play an even more
important role. Here we show that for quasi-one-dimensional structures such
as nanotubes and nanowires, side contact with the metal only leads to weak
band re-alignement, in contrast to bulk metal-semiconductor contacts.
Schottky barriers are much reduced compared with the bulk limit, and should
facilitate the formation of good contacts. However, the conventional strategy 
of heavily doping
the semiconductor to obtain ohmic contacts breaks down as the nanowire
diameter is reduced. The
issue of Fermi level pinning is also discussed, and it is demonstrated that
the unique density of states of quasi-one-dimensional structures make them
less sensitive to this effect. Our results agree with recent experimental work, and
should apply to a broad range of quasi-one-dimensional materials.
\end{abstract}

\begin{multicols}{2}

The early work of Schottky, Mott and Bardeen has laid the course for much of
the fundamental understanding and improvement in the performance of
electrical contacts to bulk semiconductors. For nanoelectronics, contacts
are a significant fraction of the total device size and are expected to be
crucial for device behavior. However, because of the unique properties of
nanostructures, much of the basic fundamental aspects of contacts need to be
re-examined at the nanoscale.

The current understanding of contacts to nanostructures is in its infancy,
both from an experimental and theoretical perspective. An example is carbon
nanotubes (NTs): despite much experimental work, it is still unclear wether
the contacts are Schottky or ohmic, with reports of Schottky contacts for Ti%
\cite{ibm1} and ohmic contacts for Au\cite{mceuen} and Pd\cite{dai1,dai2}.
However, recent experimental work\cite{chen,kim} has suggested that the type
of contact depends on the NT, with Schottky contacts for small diameter NTs
and ohmic contacs for large diameter NTs.

From a theoretical perspective, it has been demonstrated that the concept of
Fermi level pinning (crucial in traditional semiconductors) is ineffective
for quasi-one-dimensional nanostructures {\it end-bonded} to metals\cite%
{leonard}. For NTs {\it side-contacted} by a metal, modeling has been used
to extract Schottky barriers from experimental measurements of the ON
current in NT transistors\cite{chen}, but have not addressed the origin of
the Schottky barriers; and atomistic calculations have provided case-by-case
studies\cite{xue,shan} for {\it planar} contacts. However, a more general
theoretical understanding for side-contacts to quasi-one-dimensional (Q1D)
structures is still missing, especially in light of the recent experimental
findings.

In this paper, we present a theoretical and modeling analysis of side
contacts to nanotubes and nanowires. We show that the conventional concepts
developed for bulk metal-semiconductor contacts do not simply carry over to
the nanoscale. In particular, band re-alignement due to charge transfer is
weak due to the limited available depletion width. In NTs, this leads to
relatively small and slowly varying Schottly barriers with NT diameter. In
nanowires (NWs), there is a range of diameters with minimized Schottky
barriers, providing optimal contact properties. We also demonstrate that in
general, Q1D structures are much less sensitive to Fermi level pinning than
their bulk counterparts. Finally, a conventional strategy for making ohmic
contacts is to heavily dope the semiconductor near the contact; we show that
at typical dopings for Si, the contact resistance increases rapidly as the
nanowire diameter is decreased below 40 nm.

\begin{figure}[h]
\psfig{file=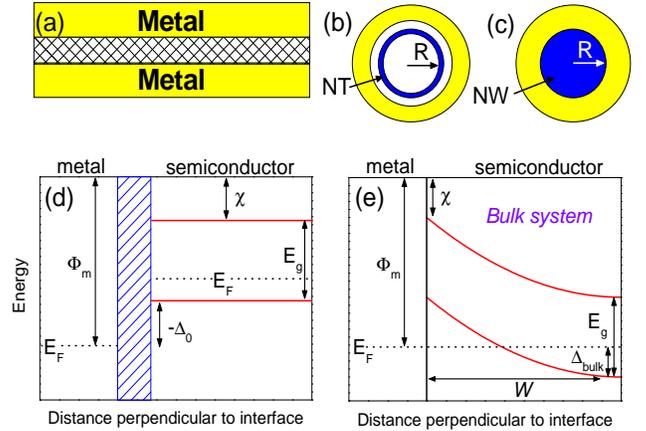,height=170pt,width=240pt}
\caption{Panel (a) shows a cross-section of the metal-nanostructure contact
along the length of the nanostructure. Panels (b) and (c) show radial
cross-sections for metal-nanotube and metal-nanowire contacts. The NT is
separated from the metal by a distance $s=0.3$ nm. The NW is modeled as a
solid cylinder with a sharp interface with the metal. Panel (d) shows the
band alignment at a metal-semiconductor contact before charge transfer. In a
bulk contact, panel (e), band bending over a distance $W$ due to charge
transfer leads to a Schottky barrier $\Delta _{bulk\text{.}} $}
\end{figure}

We begin by describing the contact geometry considered here. Figure 1a shows
a sketch of a cross section of the contact consisting of a Q1D structure
embedded in a metal. For explicit systems, we consider a single-wall NT , as
shown in Fig. 1b, or a solid nanowire as in Fig. 1c. For the NT, the metal
forms a cylindrical cavity of radius $R+s$ where $R$ is the NT radius and $s$
is the distance between the NT and the metal, while for the NW we consider a
solid, continuum cylinder embedded in a perfect metal, with a sharp
interface between the nanowire surface and the metal.

In the simplest picture, the difference between the metal Fermi level $E_{F}$
and the semiconductor valence band edge $E_{v}$ (the barrier for holes)  is
simply given by (Fig. 1d)%
\begin{equation}
\Delta _{0}=E_{F}-E_{v}=\frac{1}{2}E_{g}-\Phi _{m}+\Phi _{s}  \label{bare}
\end{equation}%
where $\Phi _{m}$ and $\Phi _{s}$ are the metal and semiconductor
workfunctions respectively, and $E_{g}$ is the semiconductor band gap. A
positive value for $\Delta _{0}$ indicates a Schottky barrier, while a
negative value indicates an ohmic contact. Because bandgap decreases with
increasing diameter for Q1D structures, the value of $\Delta _{0}$ depends
on the nanostructure diameter. The behavior of Eq. $\left( \ref{bare}\right) 
$ for undoped NTs is shown in Fig. 2 for a value of $\Phi _{m}-\Phi _{NT}=0.4
$ eV (typical of Pd), and using the relation $E_{g}=2a\gamma /d$ between
bandgap and NT diameter $d$ ($a=0.142$ nm is the C-C bond length and $\gamma
=2.5$ eV is the tight-binding overlap integral). One problem with this
picture (besides the fact that the physics is incomplete, as will be
discussed below) is that Eq. $\left( \ref{bare}\right) $ predicts large and
negative values for $\Delta _{0},$ signaling strong ohmic contacts. However,
it is clear that such strong ohmic contacts are not observed experimentally.

In general, because the metal Fermi levels of the metal and semiconductor
are not equal, charge transfer between the metal and semiconductor occurs,
and leads to band re-alignement. At a bulk semiconductor junction (Fig. 1e)
this charge transfer leads to the Schottky barrier%
\begin{equation}
\Delta _{bulk}=E_{g}+\chi -\Phi _{s}  \label{bulk}
\end{equation}%
where $\chi $ is the semiconductor electron affinity. This relationship
arises because, in the bulk system, a depletion width perpendicular to the
metal-semiconductor interface is created in the semiconductor until the band
lineup in Eq. $\left( \ref{bulk}\right) $ is obtained. However for Q1D
structures, the depletion width depends {\it exponentially} on the doping%
\cite{leonard2} and is much longer than the device size for non-degenerate
doping, leading to slowly varying bands outside of the contact; and for a
three-terminal device the band-bending in the channel is governed by the
gate voltage. In either case, the band alignment is determined by that in
the contact. But for a side-contacted Q1D structure, the semiconductor is
only a few nanometers thick in the direction perpendicular to the
metal-semiconductor interface; thus only a region of the order of the
nanostructure cross-section can be depleted, giving partial band
re-alignement. The value of $\Delta $ will then be somewhere between $\Delta
_{0}$ and $\Delta _{bulk}$ (for an undoped NT or NW, $\Delta _{bulk}=E_{g}/2,
$ which would always give relatively high Schottky barriers).

Nanotubes are an extreme example of this situation, since the possible
``depletion width'' is the size of the NT wall; the charge transfer and
image charge in the metal create two nested hollow cylinders with opposite
charge, and an associated electrostatic potential. This electrostatic
potential in turn shifts the bands, and changes the amount of transferred
charge. Thus, the charge and potential must be determined self-consistently.
We can capture this behavior using analytical models for the charge and
potential. The charge per unit area on the NT can be expressed as%
\begin{equation}
\sigma =eN\int D_{NT}(E)f(E-E_{F})dE  \label{sigma}
\end{equation}%
where%
\begin{equation}
D_{NT}(E)=\frac{a\sqrt{3}}{\pi ^{2}R\gamma }\frac{\left| E+eV_{NT}\right| }{%
\sqrt{\left( E+eV_{NT}\right) ^{2}-\left( E_{g}/2\right) ^{2}}}
\end{equation}%
is the NT density of states\cite{white}, $f\left( E-E_{F}\right) $ is the
Fermi function, and $N=4/(3\sqrt{3}a^{2})$ is the atomic areal density. We
assume a uniform and sharp distribution of the charge on the NT.

For the geometry of Fig. 1, solution of Poisson's equation gives the
potential on the NT as%
\begin{equation}
eV_{NT}=-\sigma \frac{eR}{\varepsilon _{0}}\ln \frac{R+s}{R}=-\frac{e^{2}}{C}%
\sigma  \label{potential}
\end{equation}%
where $\varepsilon _{0}$ is the permittivity of free space and $C$ is the
capacitance per unit area between the metal and the NT. Equations $\left( %
\ref{sigma}\right) $ and $\left( \ref{potential}\right) $ can be solved
self-consistently for a given NT. In this model the electrostatic potential
induced on the NT modifies the barrier to%
\begin{equation}
\Delta =\Delta _{0}-eV_{NT}.
\end{equation}%
Figure 2 shows results of such calculations for parameters typical of Pd.
Clearly, the behavior is different from the simple expressions in Eqs $%
\left( \ref{bare}\right) $ and $\left( \ref{bulk}\right) $. The results
suggest that there is a transition between Schottky and ohmic behavior at a
NT diameter around 1.4 nm, in agreement with recently published experimental
data for Pd contacts\cite{chen,kim}.

The results of these calculations can be verified using an atomistic
treatment of the NT. For selected zigzag NTs, we apply a tight-binding
Green's function formalism, dividing the system into semi-infinite left and
right leads connecting a central scattering region (Fig. 1a). In the central
region, we calculate the Green's function from

\begin{equation}
G^{R}=\left[ \left( E-eV\right) I-H_{0}-\Sigma ^{R}\right] ^{-1}\text{,}
\end{equation}%
where $H_{0}$ is the tight-binding Hamiltonian for the isolated NT and $V$
is the electrostatic potential on each ring of the zigzag NT. The function $%
\Sigma ^{R}$ represents the interaction of the scattering region with the
semi-infinite NT leads. To obtain the charge density, we assume a uniform
distribution of the charge in the azimuthal direction, and spatially
distribute the total charge on each ring in the radial and axial directions
with a Gaussion smearing function. Thus the three-dimensional charge density
is given by 
\begin{equation}
\rho (r,\phi ,z)=-\sum_{l}g(z-z_{l},r-R)\frac{e}{\pi }\int dE%
\mathop{\rm Im}%
G_{ll}^{R}  \label{charge}
\end{equation}%
where $g(x,y)$ $=\left( 4\pi ^{2}R\sigma _{z}\sigma _{r}\right) ^{-1}\exp %
\left[ -x^{2}/2\sigma _{z}^{2}-y^{2}/2\sigma _{r}^{2}\right] $ with $R$ the
tube radius, $z_{l}$ the position of ring $l$, and $\sigma _{z}$ and $\sigma
_{r}$ the smearing lengths in the axial and radial directions respectively
(this expression for $g$ is valid when $R\gg \sigma _{r}$, and we use values
of $\sigma _{z}=0.14$ nm and $\sigma _{r}=0.06$ nm).

The electrostatic potential is calculated by solving Poisson's equation in
cylindrical coordinates on a grid with the geometry of Fig. 1. Once the
three-dimensional electrostatic potential is obtained, the value for $V$ on
each layer is taken as the value of the three-dimensional electrostatic
potential at the atomic position of each ring along the NT. We iterate the
calculation of the charge and the electrostatic potential until
self-consistency.

As shown in Fig. 2, the results of such calculations indicate excellent
agreement with the analytical approach introduced above. This agreement
allows us to use equations $\left( \ref{sigma}\right) $ and $\left( \ref%
{potential}\right) $ to gain further understanding of the properties of
these contacts, as we now discuss.

\begin{figure}[h]
\psfig{file=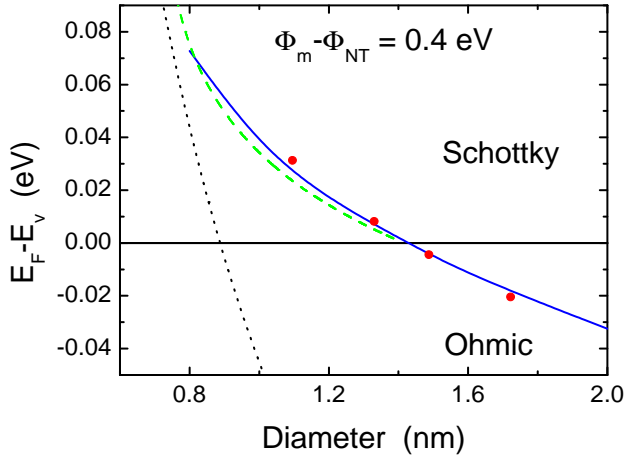,height=180pt,width=240pt}
\caption{Schottky barrier at nanotube-metal contacts for parameters typical
of Pd. Dotted line is prediction from Eq. (\ref{bare}), solid line is
computed from Eqs (\ref{sigma}) and (\ref{potential}), dashed line is Eq. (%
\ref{delta}) and circles are calculated from atomistic approach.}
\end{figure}

To proceed further we focus on the small and positive $\Delta $ regime, and
approximate the fermi distribution as $f\left( E-E_{F}\right) \approx \frac{1%
}{2}\exp \left[ -\beta \left( E-E_{F}\right) /kT\right] $ where $\beta =0.7$%
. Asymptotic expansion of the integral in Eq. \ref{sigma} leads to the charge%
\begin{equation}
\sigma =\frac{eNa\sqrt{3}}{2\sqrt{2\beta }\pi ^{3/2}R\gamma }\sqrt{\frac{%
E_{g}kT}{2}}e^{-\beta \frac{\Delta }{kT}}.  \label{chargeanalytical}
\end{equation}%
Combined with Eq. $\left( \ref{potential}\right) $ this expression for the
charge gives the Schottky barrier%
\begin{equation}
\Delta =\Delta _{0}+\frac{kT}{\beta }L\left( \beta \alpha \sqrt{\frac{E_{g}}{%
2kT}}e^{-\beta \frac{\Delta }{kT}}\right)
\end{equation}%
where $\alpha =\left( e^{2}Na\sqrt{3}\right) /\left( 2\sqrt{2\beta }\pi
^{3/2}R\gamma C\right) $ and $L(x)$ is Lambert's W function. A more
appealing formula can be obtained using asymptotic expansion of $L(x)$ giving%
\begin{equation}
\Delta =\frac{kT}{\beta }\ln \left( \frac{\alpha \sqrt{\frac{E_{g}}{2kT}}}{%
\ln \alpha \sqrt{\frac{E_{g}}{2kT}}-\Delta _{0}/kT}\right) .  \label{delta}
\end{equation}%
The behavior of this function is plotted in Fig. 2, showing good agreement
with the full calculation. The logarithmic dependence implies relatively
slowly varying $\Delta ,$ at least compared with Eq. $\left( \ref{bare}%
\right) $. The NT diameter delimiting Schottky from ohmic behavior is\cite%
{remark} 
\begin{equation}
d\approx d_{0}\left( 1+\alpha \sqrt{\frac{kT}{\Phi _{m}-\Phi _{NT}}}\right) ;
\end{equation}%
thus the diameter is increased from its bare value by $\delta d=\alpha \sqrt{%
\frac{kT}{\Phi _{m}-\Phi _{NT}}} d_{0} $. Making ohmic contact to a wide
range of NT diameters requires a small $\delta d$; this can be accomplished
at low temperature, or with a large metal workfunction. Because $\alpha $ is
inversely proportional to the capacitance, our result also provides one
reason why embedding the NT in the metal and heating can improve the contact
properties: the embedded contact provides a larger capacitance compared to a
planar contact (for typical parameters we find $\alpha \approx 2.5$ for the
embedded contact and up to $\alpha \approx 14$ for a planar contact), while
heating presumably provides a smaller average value for $s$ and a smaller
capacitance. The large value of $\alpha $ for planar contacts imply that
this type of contact is almost pathological, since the value of $\delta d$
should be very large unless the metal worfunction itself is large. This may
explain why ab initio calculations find Schottky barriers\cite{xue,shan}
despite the large values of $\Phi _{m}-\Phi _{NT}$ used in the calculations.
It is worth mentioning that changing the contact geometry to modify the
contact properties is not possible in bulk contacts, and is thus a unique
feature of nanostructures.

We now consider side-contacts to nanowires. The difference between NWs and
NTs is two-fold: first, in NWs the possible depletion width increases with
diameter, while in a NT it stays constant; second, while the band gap also
decreases with diameter for NWs\cite{SiNW}, the large diameter limit can
lead to a finite band gap, while for NTs it leads to a zero band gap. It is
not clear a priori how these effects influence the contact behavior. To
address this issue, we consider a model NW with density of states%
\begin{equation}
D_{NW}(E)=\frac{\sqrt{2m^{\ast }}}{\pi \hbar }\left( E-E_{g}/2\right) ^{-1/2}
\label{dosNW}
\end{equation}%
where $m^{\ast }$ is the effective mass. For silicon NWs, it has been shown
experimentally\cite{SiNW} that the band gap depends on diameter as $%
E_{g}=E_{0}+C/d^{2}$ where $E_{0}=1.12$ eV and $C=4.33$ eVnm$^{2}$. We
consider the situation $\Phi _{m}-\Phi _{NW}=0.7$ eV typical of contacts to
Si. Fig. 3a shows the expected Schottky barrier heights from Eq. $\left( \ref%
{bare}\right) ,$ which predicts ohmic contacts to NWs with diameters larger
than 4 nm. To study the effects of charge transfer, we perform a
self-consistent calculation of the charge and potential, using Eq. $\left( %
\ref{dosNW}\right) $ to obtain the charge and solving Poisson's equation
numerically in the NW to obtain the potential (we use an atomic volume
density $N_{v}=5\times 10^{28}$ atoms/m$^{3}$). 

\begin{figure}[h]
\psfig{file=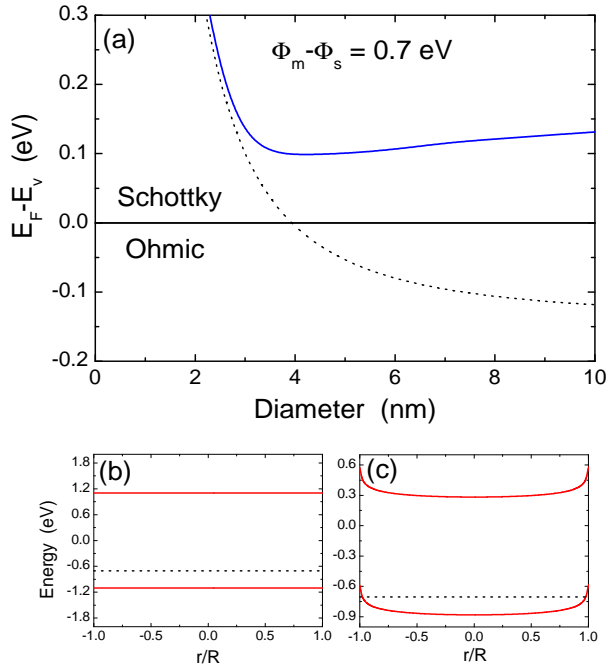,height=270pt,width=240pt}
\caption{Panel (a): Schottky barrier at nanowire-metal contacts for
parameters typical of SiNWs. Dotted line is prediction from Eq. (\ref{bare})
and solid line is self-consistent calculation. Panels (b) and (c):
Band-bending across nanowires with diameters of 2 nm and 10 nm,
respectively; dotted lines are Fermi level.}
\end{figure}

Fig. 3b,c shows the
self-consistent band-bending for NWs of 2 and 10 nm radius. Clearly, the
nanoscale dimension of the NWs prevents the bands from reaching their
asymptotic value; instead, there is only a relatively weak band-bending
present. To quantify the Schottky barrier height, we calculate the spatial
average of $E_{F}-E_{v}(r)$; the results plotted in Fig. 3a indicate a
remarkable behavior: the contact is always of Schottky character, with the
barrier minimized at a diameter of about 4 nm. Thus, while in NTs the
barrier height decreases monotonically with diameter, the behavior in other
Q1D structures may be non-monotonic, with a range of diameters providing
optimal contact properties.

We now discuss the issue of Fermi level pinning. As shown in Fig. 4a, in a
bulk metal-semiconductor contact, metal-induced gap states (MIGS) lead to
band-bending over a distance $l\ll W$, and modification of the Schottky
barrier height to $\Delta _{pin}$. The question is to what extent this
mechanism influences the properties of side contacts to NTs and NWs. To
model this effect, we consider a radial pinning charge\cite{leonard}

\begin{equation}
\sigma _{pin}(r)=D_{0}N_{A}\left[ E_{F}-E_{N}(r)\right] h(r)
\end{equation}%
where the neutrality level $E_{N}$ is at midgap [i.e. $E_{N}(r)=-eV(r)$], $%
h(r)=e^{-r/l}$ for a NW and $h(r)=\delta _{r,R}$ for a NT, and $N_{A}=N$ for
a NT and $N_{A}=N_{v}^{2/3}$ for a NW. We choose $l=0.3$ nm, a typical value
for metal-semiconductor interfaces\cite{louie}. We add this pinning charge
to Eq. $\left( \ref{sigma}\right) $ or to the charge calculated from Eq. $%
\left( \ref{dosNW}\right) $ and repeat our self-consistent calculations.

Figure 4b shows the Schottky barrier calculated for several NTs as a
function of the density of gap states ($\Delta_{pin}=E_{g}/2$). Clearly,
there is a rapid onset of pinning at $D_{0}\sim 0.1$ states/(atom$\cdot $%
eV); this value of $D_{0}$ is rather larger considering the van der Waals
bonding of NTs to metals. Furthermore, atomistic calculations\cite{xue} have
obtained values at least an order of magnitude smaller. Thus, as in
end-bonded contacts, we expect that Fermi level pinning will play a minor
role in side-contacts to NTs. The rather large value of $D_{0}$ required to
see pinning effect is a generic property of Q1D structures as we now discuss.

Figure 4c shows the effects of Fermi level pinning on the barrier height in
SiNWs. The results also indicate a value of $D_{0}\sim 0.1$ states/(atom$%
\cdot $eV) required to see pinning effects. For comparison, the inset in
this figure shows the same calculation for a bulk metal-semiconductor
interface with the same parameters, indicating that only $0.002$ states/(atom%
$\cdot $eV) are needed to reach the onset of pinning. Thus, the Q1D system
requires almost two orders of magnitude larger density of pinning states
compared with the bulk interface.

The origin of this behavior can be traced to the unique density of states of
Q1D systems. Indeed, for Si, we can repeat the analysis leading to Eq. $%
\left( \ref{chargeanalytical}\right) $ using the density of states for the
NW and for the bulk system [$D_{bulk}(E)=\sqrt{2}\left( m^{\ast }\right)
^{3/2}\left( \pi ^{2}\hbar ^{3}\right) ^{-1}\sqrt{E-E_{g}/2}$]. This leads
to the ratio $\sigma _{NW}/\sigma _{bulk}=\left( 2\pi N_{v}^{1/3}\beta
\right) /\left( m^{\ast }kT\right) $. 

\begin{figure}[h]
\psfig{file=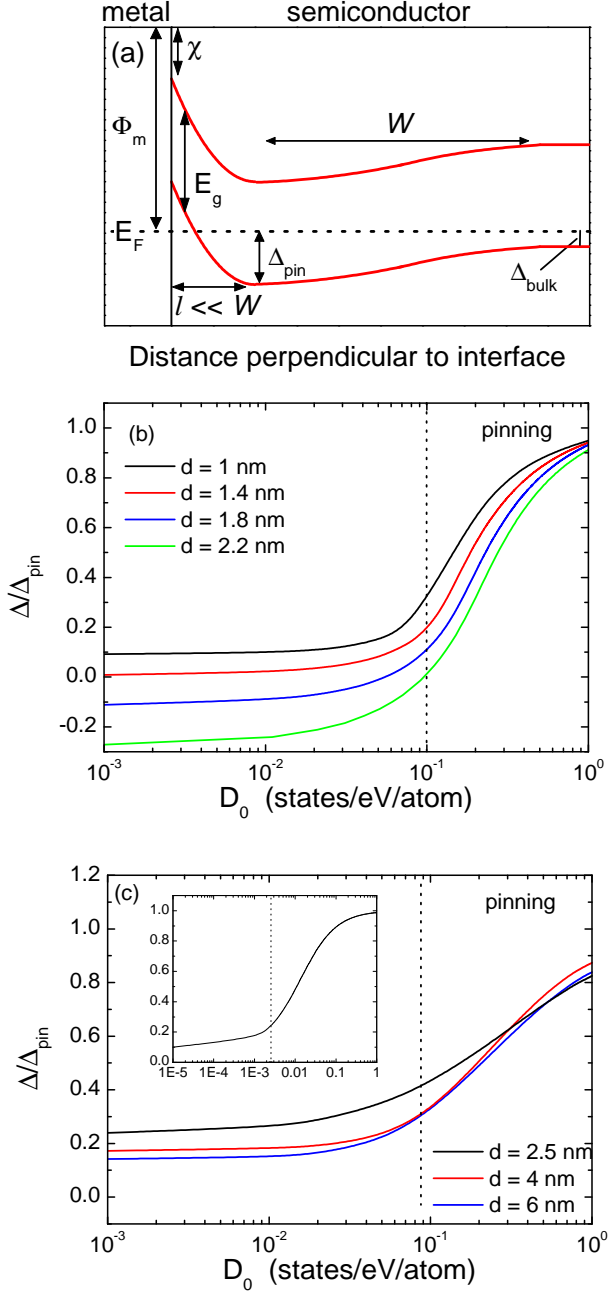,height=500pt,width=240pt}
\caption{Panel (a) shows a sketch of the band-bending in the presence of
metal-induced gap states at a bulk metal-semiconductor interface. Panels (b)
and (c) show the calculated Schottky barrier as a function of the density of
gap states for several NTs and NWs, respectively. The inset in (c) shows the
behavior for a planar metal-semiconductor contact.}
\end{figure}

The appearance of the $kT$ factor in
the denominator is entirely due to the Q1D density of states of the NW and
the presence of a van Hove singularity at the band edge. At room
temperature, we find that $\sigma _{NW}/\sigma _{bulk}>100$; thus the MIGS
are competing with a much larger charge density in the Q1D system.

\begin{figure}[h]
\psfig{file=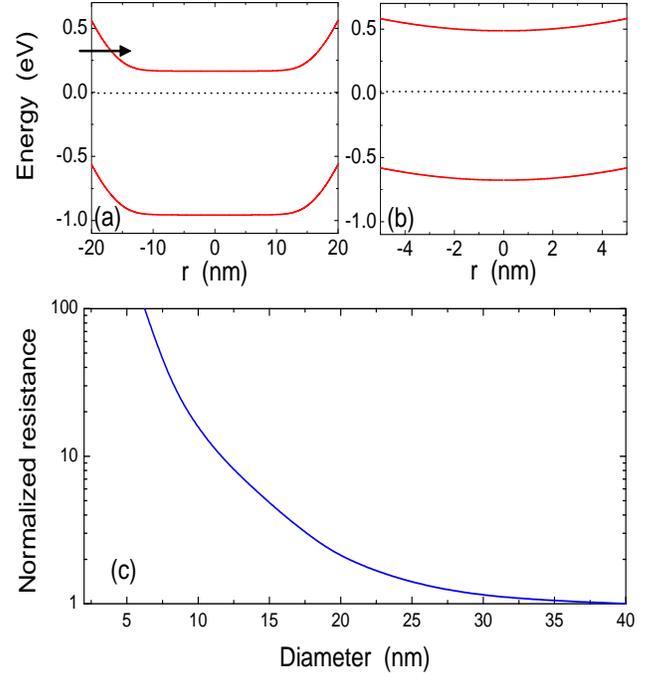,height=270pt,width=240pt}
\caption{Band-bending across Si NWs with doping of 10$^{19}e/cm^{-3}$ for
diameters of 40 nm (a) and 10 nm (b). The arrow indicates tunneling of
electrons throught the Schottky barrier. The normalized resistance is shown
in figure (c) as a function of NW diameter.}
\end{figure}

Our discussion has so far focused on the situation of low doping, where the
strategy for making ohmic contacts is by selection of a metal with
appropriate workfunction. In traditional metal-semiconductor contacts, an
alternative approach is to heavily dope the semiconductor, and rely on
tunneling through the Schottky barrier to reduce the contact resistance and
obtain ohmic-like contatcs. To address the feasibility of this approach for
contacts to NWs, we repeat our self-consistent calculations for the Si NW,
focusing on the situation where the metal Fermi level is in the middle of
the NW bandgap at the interface, and adding a uniform doping charge of $%
10^{19}e/cm^{-3}$. Figure 5 shows the band-bending in the presence of this
doping charge for NWs of 40 and 10 nm diameters. We calculate the contact
conductance from%
\begin{equation}
G\sim \int_{E_{c}^{\min }}^{\infty }T(E)\left( -\frac{\partial f}{\partial E}%
\right) dE
\end{equation}%
where the tunneling probability $T(E)$ is obtained from the WKB
approximation. The normalized contact resistance is then $G_{\infty }/G$
where $G_{\infty }$ is the conductance in the limit of large diameters. The
behavior of the normalized resistance as a function of NW diameter is shown
in Fig. 5c, indicating a rapid increase of the resistance with decrease in
diameter. The origin of this behavior is due to the increased tunneling
distance and reduced range of tunneling energies because of the poor
band-bending in the NW (the increase in bandgap also plays a role for the
smaller diameters). One implication of this result is that different
diameter NWs will require different doping levels to achieve the same
contact quality.

In summary, we find that the concepts developed to describe traditional
metal-semiconductor interfaces fail to properly account for the properties
of contacts to quasi-one-dimensional structures such as nanowires and
nanotubes. The nanometer cross-section of nanowires and nanotubes prevents
the formation of an appropriate depletion width, leading to only weak band
re-alignement; and because of the diverging density of states at the band
edge, these nanostructures are much less sensitive to Fermi level pinning
than their bulk counterparts. Optimizing device performance will not only
require selecting Q1D structures for their behavior in the channel, but also
for their contact properties. We expect that our results will be applicable
to a broad range of Q1D structures.

\bigskip $^{\ast }$email:fleonar@sandia.gov

{\bf Acknowledgement. }Sandia is a multiprogram laboratory operated by
Sandia Corporation, a Lockheed Martin Company, for the United States
Department of Energy under contract DE-AC01-94-AL85000.

\end{multicols}


\begin{references}
\bibitem{remark} To derive this relation we assumed a constant capacitance
over the range of NT diameters of interest. A more detailed treatment would
not change the key physics.
\end{references}
\end{document}